\documentclass[authoryear,final,5p,twocolumn,times]{elsarticle}
\usepackage{amsmath,amssymb,bm,color,graphicx,hyperref} % setspace,mathptmx

\DeclareMathOperator{\Real}{Re}

\newcommand{\ri}{\mathrm{i}}
\newcommand{\rd}{\mathrm{d}}
\newcommand{\re}{\mathrm{e}}

\journal{Mechanics Research Communications}

\begin{document}

\begin{frontmatter}

\title{On a difficulty in the formulation of initial and boundary conditions for eigenfunction expansion solutions for the start-up of fluid flow} %of Newtonian and non-Newtonian fluids}

\author{Ivan C. Christov}
\ead{christov@alum.mit.edu}
\ead[url]{http://alum.mit.edu/www/christov}
\address{Department of Mechanical and Aerospace Engineering, Princeton University, Princeton, NJ 08544, USA}

%\date{\today}

\begin{abstract}
Most mathematics and engineering textbooks describe the process of ``subtracting off'' the steady state of a linear parabolic partial differential equation as a technique for obtaining a boundary-value problem with homogeneous boundary conditions that can be solved by separation of variables (i.e., eigenfunction expansions). While this method produces the correct solution for the start-up of the flow of, e.g., a Newtonian fluid between parallel plates, it can lead to erroneous solutions to the corresponding problem for a class of non-Newtonian fluids. We show that the reason for this is the non-rigorous enforcement of the start-up condition in the textbook approach, which leads to a violation of the principle of causality. Nevertheless, these boundary-value problems can be solved correctly using eigenfunction expansions, and we present the formulation that makes this possible (in essence, an application of Duhamel's principle). The solutions obtained by this new approach are shown to agree identically with those obtained by using the Laplace transform in time only, a technique that  enforces the proper start-up condition implicitly (hence, the same error cannot be committed).
\end{abstract}


\end{frontmatter}

% ----------------------------------------------------------------
\section{Introduction: Stokes' first problem on a strip (start-up of plane Couette flow)}
\label{sec:intro}

Suppose an incompressible Newtonian fluid with kinematic viscosity $\nu$ fills the gap between two parallel horizontal plates a distance $d$ apart. We are interested in the problem of the bottom plate suddenly being set into motion at $t=0^+$ with constant velocity $U_0$ in the positive $x$-direction. Known as plane Couette flow, this is a variation on a problem first considered by Sir George Gabriel Stokes \citep{S51}, nowadays called ``{Stokes' first problem},'' which has applications in rheometry and creep tests of deformation \citep{Netal07}. Stokes-type problems remain an important aspect of modern non-Newtonian fluid mechanics \citep{DMO08,PMW11,MPW12,VJVHF13}. Choosing the dimensionless variables $u = u^\star/U_0$, $y=y^\star/d$ and $t=t^\star/(d^2/\nu)$, we must solve an initial-boundary-value problem (IBVP) for the $x$-component of the velocity field, $u = u(y,t)$, of the fluid \citep[see, e.g.,][]{B67,L07}:
\begin{subequations}\begin{align}
\frac{\partial u}{\partial t} &= \frac{\partial^2 u}{\partial y^2},\qquad (y,t) \in (0,1)\times(0,\infty),\label{eq:newt_pde}\displaybreak[3]\\
u(y,0) &= 0,\qquad 0<y<1,\displaybreak[3]\\
u(0,t) &= 1,\qquad t > 0,\label{eq:newt_bc}\\
u(1,t) &= 0,\qquad t > 0.
\end{align}\label{eq:newt}\end{subequations}

In what follow we shall contrast the latter IBVP with an example IBVP corresponding to the plane Couette flow of an incompressible rate-type non-Newtonian fluid, the rheology of which we take to have the dimensionless form 
\begin{equation}
\sigma + \tau \frac{\partial \sigma}{\partial t} = \dot\gamma + \alpha \frac{\partial \dot\gamma}{\partial t},
\label{eq:nonnewt_rheol}
\end{equation}
where $\sigma$ is the shear stress, $\dot\gamma = \partial u/\partial y$ is the rate of strain, $\tau = \lambda_1/(d^2/\nu)$ is the dimensionless \emph{relaxation time}, and $\alpha = \lambda_2/(d^2/\nu)$ is the dimensionless \emph{retardation time} with $0 < \lambda_2 < \lambda_1$ in order to have a causal and well-posed model \citep[see, e.g.,][Sect.~5.2b]{BAH87}. Now, eliminating $\sigma$ between Eq.~\eqref{eq:nonnewt_rheol} and Cauchy's first law $\partial u/\partial t = \partial\sigma/\partial y$ (see, e.g., \citep{T62a,C10}), the non-Newtonian version of IBVP~\eqref{eq:newt} that we shall consider takes the form
\begin{subequations}\begin{align}
\left(1 + \tau\frac{\partial}{\partial t}\right)\frac{\partial u}{\partial t} &= \frac{\partial}{\partial y} \left(1 + \alpha\frac{\partial}{\partial t}\right)\frac{\partial u}{\partial y},\qquad (y,t) \in (0,1)\times(0,\infty), \label{eq:nonnewt_pde}\\
u(y,0) &= \frac{\partial u}{\partial t}(y,0) = 0,\qquad 0<y<1,\\
u(0,t) &= 1,\qquad t > 0,\label{eq:nonnewt_bc}\\
u(1,t) &= 0,\qquad t > 0.
\end{align}\label{eq:nonnewt}\end{subequations}
To the best of our knowledge, the correct solution to IBVP~\eqref{eq:nonnewt} cannot be found in the literature.

Equation~\eqref{eq:nonnewt_rheol} is attributed \citep{J86,BAH87} to Sir Harold Jeffreys \citep{J29,J32}, and it appears to have been independently proposed as a constitutive relation for polymeric suspensions \citep{FS46}. This model has been successfully applied to experimental data \citep{TS53}, it has been generalized in a frame-indifferent manner to multiple dimensions by \citet{O50}, and more recently it was extended to apply to bubbly liquids as well \citep{LMW02}. Jeffreys' rheology\footnote{Though many recent papers refer to Eq.~\eqref{eq:nonnewt_pde} as corresponding to an ``Oldroyd-B fluid,'' it should be noted that the contribution of Oldroyd is in the nonlinear terms of the rheological model, which are self-canceling for planar unidirectional flows. Therefore, there is no sense in which to distinguish between Oldroyd's upper-convected (``B'') and lower-convected (``A'') time rates, and it is more appropriate to credit this equation to Jeffreys, as done in the textbook of \citet{BAH87}.} covers the well-known relaxation model due to James Clerk Maxwell \citep{M67} as special case in the limit of $\alpha\to0$. Meanwhile the rheology of the so-called second-order (or second-grade) fluid \citep[Sect.~7]{CN60} undergoing planar unidirectional flows is recovered in the limit of $\tau\to0$. Thus, for our purposes, IBVP~\eqref{eq:nonnewt} is a sufficiently general non-Newtonian start-up problem. Finally, note that Eq.~\eqref{eq:nonnewt_pde} is also a generic model for the evolution of linearly dissipative-dispersive wave packets \citep{C08}.

Though it is standard practice \citep{CJ59,B67,L07} to write the boundary conditions as done in Eqs.~\eqref{eq:newt_bc} and \eqref{eq:nonnewt_bc}, note that this leaves a piece of the problem somewhat vague, i.e., not explicitly stated. That is, prior to start-up (i.e., for $t<0$) $u \equiv 0$. This means that the boundary condition for Stokes' first problem suffers a jump discontinuity across the plane $t=0$. Hence, the mathematically-correct way of writing it is $u(0,t) = H(t)$. The purpose of the present article is to show that this is not merely a semantic distinction of no consequence, and that it fundamentally affects the method of solution. Specifically, one cannot arrive at the correct solution of IBVP~\eqref{eq:nonnewt} using the boundary condition $u(0,t)=1$ together with the textbook techniques \citep{CJ59,B67,L07,B08}. Meanwhile, the solution to IBVP~\eqref{eq:newt} happens to be unaffected. This has significant implications not only for problems of viscoelastic fluid flow (as in the featured example) but also for problems of hyperbolic, delayed or otherwise nonclassical heat conduction, wherein similar IBVPs arise \citep{T97,S11}. Such IBVPs are also relevant in acoustics, where they are referred to as \emph{signaling problems} \citep{C98}. Historically, identifying such apparent mathematical ``difficulties'' has proven worthwhile, e.g., the theory of shock waves was developed, in part, because of Stokes' paper ``On a difficulty in the theory of sound'' \citep{S48}.

This paper is organized as follows. In Sect.~\ref{sec:ef_wrong}, the solutions of IBVPs~\eqref{eq:newt} and \eqref{eq:nonnewt} are found by the textbook eigenfunction expansion technique. In Sect.~\ref{sec:laplace}, the solutions are derived by using only the Laplace transform in time, showing that the Laplace-transform solution to IBVP~\eqref{eq:nonnewt} does not agree with the solution found in Sect.~\ref{sec:ef_wrong}. Then, in Sect.~\ref{sec:ef_correct}, the textbook eigenfunction expansion technique is modified to satisfy causality and shown to reproduce identically the (correct) solutions to both IBVPs found in Sect.~\ref{sec:laplace}. Sect.~\ref{sec:discussion} gives a critical discussion of the literature on start-up flows in the context of the present work. %{\color{red}Finally, a cocksure reviewer of this manuscript erroneously insisted that the present paradox is resolved by applying the textbook technique to the \emph{system} of PDEs for $\{u,\sigma\}$, instead of the single-PDE formulation in \eqref{eq:nonnewt_pde}. Thus, it is shown in Appendix~\ref{app:system} that this leads to yet another wrong solution.}

% ----------------------------------------------------------------
\section{Solution by separation of variables through a transformation to homogeneous boundary conditions}
\label{sec:ef_wrong}

If we assume $u$ is independent of $t$, then we obtain the steady-state solution of both IBVPs \eqref{eq:newt} and \eqref{eq:nonnewt}, namely
\begin{equation}
u_\mathrm{ss}(y) = 1-y.
\label{eq:ss}
\end{equation}
Using this observation, the textbook approach \citep{CJ59,B67,L07,B08} is to now make the change of dependent variable
\begin{equation}
u(y,t) = v(y,t) + u_\mathrm{ss}(y),
\label{eq:subst_ss}
\end{equation}
where $v(0,t) = v(1,t) = 0$, unlike $u(y,t)$. Noting that $\partial u_\mathrm{ss}/\partial t = \partial^2 u_\mathrm{ss}/\partial y^2 = 0$, IBVP~\eqref{eq:newt} becomes
\begin{subequations}\begin{align}
\frac{\partial v}{\partial t} &= \frac{\partial^2 v}{\partial y^2},\qquad (y,t) \in (0,1)\times(0,\infty),\label{eq:newt_subs_ss_pde}\displaybreak[3]\\
v(y,0) &= -u_\mathrm{ss}(y),\qquad 0<y<1,\\
v(0,t) &= 0,\qquad t > 0,\\
v(1,t) &= 0,\qquad t > 0.
\end{align}\label{eq:newt_subts_ss}\end{subequations}
Similarly, IBVP~\eqref{eq:nonnewt} becomes
\begin{subequations}\begin{align}
\left(1 + \tau\frac{\partial}{\partial t}\right)\frac{\partial v}{\partial t} &= \frac{\partial}{\partial y} \left(1 + \alpha\frac{\partial}{\partial t}\right)\frac{\partial v}{\partial y},\qquad (y,t) \in (0,1)\times(0,\infty),\label{eq:nonnewt_subs_ss_pde}\displaybreak[3]\\
v(y,0) &= -u_\mathrm{ss}(y),\quad \frac{\partial v}{\partial t}(y,0) = 0,\qquad 0<y<1,\label{eq:nonnewt_ef_ic}\\
v(0,t) &= 0,\qquad t > 0,\\
v(1,t) &= 0,\qquad t > 0.
\end{align}\label{eq:nonnewt_subst_ss}\end{subequations}

It is well known \citep{T62} that the eigenvalue problem
\begin{equation}
\frac{\rd^2}{\rd y^2}\psi_n(y) = -\lambda_n\psi_n(y),\qquad \psi_n(0) = \psi_n(1) = 0
\end{equation}
possesses a complete set of orthogonal eigenfunctions $\{\psi_n(y)\}_{n=1}^{\infty}$, namely
\begin{multline}
\psi_n(y) = \sin\left(\sqrt{\lambda_n}\, y\right),\qquad \lambda_n = n^2\pi^2,\\ \int_0^1 \psi_n(y)\psi_m(y)\,\rd y = \frac{1}{2}\delta_{nm},
\label{eq:eigenfunc}
\end{multline}
where $\delta_{nm}$ is the Kronecker delta symbol and $n,m=1,2,\hdots$.

With this in mind, the method of \emph{separation of variables} \citep[see, e.g.,][Chap.~6, Sect.~4]{S06} suggests the ansatz $v(y,t) = \sum_n a_n(t)\psi_n(y)$. Substituting the latter into Eq.~\eqref{eq:newt_subs_ss_pde} and using the orthogonality relation from Eq.~\eqref{eq:eigenfunc}, we see that $a_n$ must satisfy
\begin{equation}
\frac{\rd a_n}{\rd t} = -\lambda_n a_n,
\label{eq:ode_a_new}
\end{equation}
whence $a_n(t) = a_n(0)\re^{-\lambda_n t}$ and $a_n(0) = 2\int_0^1 v(y,0)\psi_n(y) \,\rd y = -2/(n\pi)$. Thus, we would be led to believe that the solution to IBVP~\eqref{eq:newt} is
\begin{equation}
u(y,t) = (1 - y) - \frac{2}{\pi}\sum_{n=1}^\infty \exp\left(-n^2\pi^2 t\right) \frac{\sin(n\pi y)}{n}.
\label{eq:soln_ef_newt}
\end{equation}
Indeed this is precisely the dimensionless version of the expression found in Eq.~(1) (with $v_1=1$, $v_2=0$ and $f(x)=0$) in Sect.~3.4 of the book of \citet{CJ59}, in Eq.~(4.3.14) in Sect.~4.3 of the book of \citet{B67} and in Eq.~(3-158) in Chap.~3F of the book of \citet{L07}. However, notice that (due to our choice in notation) we must keep in mind that this solution does not apply for $t<0$ because prior to start-up $u \equiv 0$, which is not true for the expression in Eq.~\eqref{eq:soln_ef_newt}.

Similarly, for the Jeffreys fluid, by substituting the ansatz $v(y,t) = \sum_n a_n(t)\psi_n(y)$ into Eq.~\eqref{eq:nonnewt_subs_ss_pde} and using the orthogonality relation from Eq.~\eqref{eq:eigenfunc}, we see that $a_n$ must satisfy
\begin{equation}
\left(1 + \tau\frac{\rd}{\rd t}\right)\frac{\rd a_n}{\rd t} = -\lambda_n\left(1 + \alpha\frac{\rd}{\rd t}\right) a_n.
\label{eq:a_ode_nonnewt}
\end{equation}
The characteristic polynomial of this ordinary differential equation (ODE) is $\tau r^2 + (1 + \lambda_n\alpha) r + \lambda_n$; its roots are $r_\pm = -\left[(1+\lambda_n\alpha)\pm\sqrt{\Delta_n}\right]/(2\tau)$, where $\Delta_n := (1+\lambda_n\alpha)^2 - 4\lambda_n\tau$. Thus, three cases can be distinguished based on the sign of $\Delta_n$, and we arrive at
\begin{multline}
a_n(t) = \exp\left(-\frac{1+\lambda_n\alpha}{2\tau}t\right)\\
\times
\begin{cases}
\mathfrak{c}_n \sinh\left(\displaystyle\frac{t}{2\tau}\sqrt{\Delta_n}\right) + \mathfrak{d}_n \cosh\left(\displaystyle\frac{t}{2\tau}\sqrt{\Delta_n}\right), &\Delta_n > 0,\\
\mathfrak{c}_nt + \mathfrak{d}_n &\Delta_n = 0,\\ 
\mathfrak{c}_n \sin\left(\displaystyle\frac{t}{2\tau}\sqrt{|\Delta_n|}\right) + \mathfrak{d}_n \cos\left(\displaystyle\frac{t}{2\tau}\sqrt{|\Delta_n|}\right), &\Delta_n < 0,
\end{cases}
\end{multline}
where the constants $\mathfrak{c}_n$ and $\mathfrak{d}_n$ are determined from the initial conditions in Eq.~\eqref{eq:nonnewt_ef_ic} as follows
\begin{subequations}\begin{align}
%2\int_0^1 (y-1)\psi_n(y) \,\rd y = 
-\frac{2}{n\pi} &= a_n(0) = \mathfrak{d}_n,\\
0 &= \frac{\rd a_n}{\rd t}(0) = \begin{cases}
-\displaystyle\frac{1+\lambda_n\alpha}{2\tau}\mathfrak{d}_n + \displaystyle\frac{\sqrt{|\Delta_n|} }{2\tau}\mathfrak{c}_n, &\Delta_n \ne 0,\\[3mm]
-\displaystyle\frac{1+\lambda_n\alpha}{2\tau}\mathfrak{d}_n + \mathfrak{c}_n, &\Delta_n = 0.
\end{cases}
\end{align}\label{eq:nonnewt_ef_a_ic}\end{subequations}
Thus, we would be led to believe that the solution to IBVP~\eqref{eq:nonnewt} is
\begin{equation}
u(y,t) = (1 - y) - \frac{2}{\pi}\sum_{n=1}^\infty \exp\left(-\frac{1+\lambda_n\alpha}{2\tau}t\right) A_n(t) \frac{\sin(n\pi y)}{n},
\label{eq:soln_ef_nonnewt}
\end{equation}
where
\begin{equation}
A_n(t) = \begin{cases}
%\tau \displaystyle\frac{1-(-1)^n}{\sqrt{\Delta_n}} \sinh\left(\displaystyle\frac{t}{2\tau}\sqrt{\Delta_n}\right) - 
\displaystyle\frac{1+\lambda_n\alpha}{\sqrt{\Delta_n}}\sinh\left(\displaystyle\frac{t}{2\tau}\sqrt{\Delta_n}\right) + \cosh\left(\displaystyle\frac{t}{2\tau}\sqrt{\Delta_n}\right), &\Delta_n > 0,\\[5mm]
%\displaystyle\frac{1-(-1)^n}{2} t - 
\displaystyle\frac{1+\lambda_n\alpha}{2\tau}t + 1, &\Delta_n = 0,\\[5mm]
%\tau \displaystyle\frac{1-(-1)^n}{n\sqrt{\Delta_n}}  \sin\left(\displaystyle\frac{t}{2\tau}\sqrt{|\Delta_n|}\right) - 
\displaystyle\frac{1+\lambda_n\alpha}{\sqrt{|\Delta_n|}}\sin\left(\displaystyle\frac{t}{2\tau}\sqrt{|\Delta_n|}\right) + \cos\left(\displaystyle\frac{t}{2\tau}\sqrt{|\Delta_n|}\right), &\Delta_n < 0.
\end{cases}
\label{eq:An_wrong}
\end{equation}
Again, due to our choice in notation, we must keep in mind that this solution does not apply for $t<0$. 

Let us now find the start-up solutions by another analytical technique. If we have not made any mistakes so far, then the existence and uniqueness theorems for linear BVPs \citep{CH62} guarantee  that any solutions found by a different technique must agree with the solutions from this section.

% ----------------------------------------------------------------
\section{Solution using only the Laplace transform in time}
\label{sec:laplace}

An alternative solution technique is to apply the Laplace transform in time (see, e.g., the classic comprehensive treatment by \citet{D74}), namely 
\begin{equation}
\mathcal{L}\{u(y,t)\} := \int_0^\infty \re^{-st}u(y,t)\,\rd t \equiv \bar{u}(y,s),\quad s\in\mathbb{C}.
\end{equation}
Now, the condition of start-up, i.e., $u(0,t) \equiv 0$ for $t<0$ is implicit because the Laplace transform is only well defined for distributions whose action on tests functions with support in $t \in (-\infty,0)$ vanishes identically \citep[Sect.~12]{D74}. (This is related to the \emph{principle of causality} \citep{T56}, i.e., the Laplace transform is a causal operator only for $u(\cdot,t)$ such that $u(\cdot,t) \equiv 0$ for $t<0$.) As a consequence, $\mathcal{L}\{1\}$ \emph{must} be understood in the sense $\mathcal{L}\{H(t)\}$, where $\mathcal{L}\{H(t)\} = 1/s$, while initial conditions \emph{imposed} on $u$ must be understood in the sense of $\lim_{t\to0^-} u(y,t)$ because $\lim_{t\to0^+} u(y,t)$  does not have to attain this same value (see \citet[][pp.~70--71, 129]{D74} and \citet{MIT1803}).

Applying the Laplace transform to IBVP~\eqref{eq:newt} gives the \emph{subsidiary boundary value problem}
\begin{subequations}\begin{align}
s \bar{u} &= \frac{\partial^2 \bar{u}}{\partial y^2},\qquad y \in (0,1),\\
%u(y,0) &= \frac{\partial u}{\partial t}(y,0) = 0,\quad 0<y<1,\\
\bar{u}(0,s) &= \frac{1}{s},\\
\bar{u}(1,s) &= 0.
\end{align}\label{eq:newt_laplace}\end{subequations}
The solution of the latter BVP is easily found to be
\begin{equation}
\bar{u}(y,s) = \frac{\sinh\left[\sqrt{s}(1-y)\right]}{s \sinh(\sqrt{s})}.
\label{eq:soln_lap_dom_newt}
\end{equation}
Noting that the poles (in the complex plane) of the right-hand side are at $s$ such that $s=0$ or $\sqrt{s} = \ri n \pi$, $n=1,2,\hdots$, it is a standard exercise \citep[see, e.g.,][Sect.~40]{CJ63} to evaluate the Laplace inversion integral using a Bromwich-type contour in the complex plane and the residue theorem, arriving at 
\begin{equation}
u(y,t) = H(t)\left[ (1 - y) - \frac{2}{\pi}\sum_{n=1}^\infty \exp\left(-n^2\pi^2 t\right)\frac{\sin(n\pi y)}{n}\right].
\label{eq:soln_laplace_newt}
\end{equation}
Note the steady state is, as expected, $u_\mathrm{ss}(y) \equiv \lim_{t\to\infty} u(y,t) = 1-y$ but the change of variables given in Eq.~\eqref{eq:subst_ss} does \emph{not} eliminate this term from the solution in Eq.~\eqref{eq:soln_laplace_newt} because this change of variables does not leave the solution identically equal to zero for $t<0$, as required by the condition of start-up. We shall further elaborate on this point below.

Similarly, IBVP~\eqref{eq:nonnewt} can be transformed into
\begin{subequations}\begin{align}
\frac{s(1+\tau s)}{1+\alpha s}\bar{u} &= \frac{\partial^2 \bar{u}}{\partial y^2},\qquad y \in (0,1),\\
%u(y,0) &= \frac{\partial u}{\partial t}(y,0) = 0,\quad 0<y<1,\\
\bar{u}(0,s) &= \frac{1}{s},\\
\bar{u}(1,s) &= 0.
\end{align}\label{eq:nonnewt_laplace}\end{subequations}
The solution of the latter BVP is easily found to be
\begin{equation}
\bar{u}(y,s) = \frac{\sinh\left[\sqrt{\zeta(s)}(1-y)\right]}{s \sinh\left[\sqrt{\zeta(s)}\right]},\qquad \zeta(s) := \frac{s(1+\tau s)}{1+\alpha s}.
\label{eq:soln_lap_dom_nonnewt}
\end{equation}
The poles of $\bar{u}$ are at $s$ such that $\sqrt{\zeta(s)} = \ri n \pi$, i.e., at $s = \{0,-\tau^{-1},s_n^\pm\}$, where
\begin{equation}
s_n^\pm = \frac{-(1+ n^2\pi^2\alpha) \pm \sqrt{(1+ n^2\pi^2\alpha)^2 - 4\tau n^2\pi^2}}{2\tau},\quad n=1,2,\hdots.
\end{equation}
As before, we can use the Laplace inversion formula with a Bromwich-type contour, and all that is left to do is find the residues of $\re^{st}\bar{u}(s,t)$ at the poles (see \citep{CJ63}).\footnote{For these calculations the reader is encouraged to make use of a computer algebra system such as the software package {\sc Mathematica}.}

The pole at $s=0$ is of order 2 because both $s$ and $\sinh\left[\sqrt{\zeta(s)}\right]$ vanish, hence the residue is given by
\begin{equation}
\lim_{s\to0} \frac{\rd}{\rd s} \left\{ s^2 \frac{\re^{st} \sinh\left[\sqrt{\zeta(s)}(1-y)\right]}{s \sinh\left[\sqrt{\zeta(s)}\right]} \right\} = 1-y.
\label{eq:residue0}
\end{equation}
The pole at $s=-\tau^{-1}$ is simple and its residue is easily found to be zero. Then, since $s_n^\pm = \Big[-(1+n^2\pi^2\alpha) \pm \sqrt{\Delta_n}\Big]/(2\tau)$, where $\Delta_n = (1+ n^2\pi^2\alpha)^2 - 4\tau n^2\pi^2$, the poles at $s=s_n^\pm$ for the case $\Delta_n\ne0$ are simple and the corresponding residues are given by
\begin{equation}
\frac{\re^{s_n^\pm t} \sinh\left[\ri n \pi (1-y) \right]}{\left.\frac{\rd}{\rd s}\left( s \sinh\left[\sqrt{\zeta(s)} \right]\right)\right|_{s=s_n^\pm}} = \frac{\re^{s_n^\pm t} \sinh\left[\ri n \pi (1-y) \right]}{\left. s \cosh\left[\sqrt{\zeta(s)} \right]\frac{\zeta'(s)}{2\sqrt{\zeta(s)}}\right|_{s=s_n^\pm}}.
\label{eq:residues1}
\end{equation}
Note that $\sinh[\ri n \pi(1-y)]/\cosh(\ri n\pi) = - \ri \sin(n\pi y)$ and
\begin{equation}
\left\{\left. s\frac{\zeta'(s)}{2\sqrt{\zeta(s)}}\right|_{s=s_n^\pm}\right\}^{-1} = \frac{\mp 1 \pm  n^2 \pi^2 \alpha + \sqrt{\Delta_n}}{\ri n \pi \sqrt{\Delta_n}}.
\end{equation}
Hence, the expression on the right-hand side of Eq.~\eqref{eq:residues1} equals
\begin{equation}
-\frac{\sin(n\pi y)}{n\pi} \re^{s_n^\pm t} \left(\frac{\pm 1 \mp n^2 \pi^2 \alpha}{\sqrt{\Delta_n}} + 1\right).
\label{eq:residues2}
\end{equation}
Since Eq.~\eqref{eq:residues2} corresponds, for a given $n$, to the residues at two distinct poles, and eventually we have so add-up all residues, we can sum these now to obtain
\begin{multline}
-\frac{2\sin(n\pi y)}{n\pi}\exp\left(-\frac{1+n^2\pi^2\alpha}{2\tau} t\right)\\ \times \begin{cases} \displaystyle\frac{1 - n^2 \pi^2 \alpha}{\sqrt{\Delta_n}} \sinh\left(\displaystyle\frac{t}{2\tau}\sqrt{\Delta_n}\right) + \cosh\left(\displaystyle\frac{t}{2\tau}\sqrt{\Delta_n}\right),&\Delta_n > 0,\\[5mm] \displaystyle\frac{1 - n^2 \pi^2 \alpha}{\sqrt{|\Delta_n|}} \sin\left(\displaystyle\frac{t}{2\tau}\sqrt{|\Delta_n|}\right) + \cos\left(\displaystyle\frac{t}{2\tau}\sqrt{|\Delta_n|}\right),&\Delta_n < 0. \end{cases}
\label{eq:residuednn0}
\end{multline}
%\re^{s_n^\pm t} \left(\frac{\mp 1 \pm n^2 \pi^2 \alpha}{\Delta_n} + 1\right)%
Finally, for the case of $\Delta_n = 0$, we note that the limit $\Delta_n\to0$ is well-defined from both above and below because taking it in either expression in Eq.~\eqref{eq:residuednn0} yields
\begin{equation}
-\frac{2\sin(n\pi y)}{n\pi}\exp\left(-\frac{1+n^2\pi^2\alpha}{2\tau} t\right)\left(\frac{1-n^2\pi^2\alpha}{2\tau}t + 1\right),
\label{eq:residuedn0}
\end{equation}
hence this is the residue for $\Delta_n = 0$.

Setting $\lambda_n = n^2\pi^2$, adding the expressions in Eqs.~\eqref{eq:residue0}, \eqref{eq:residuednn0} and \eqref{eq:residuedn0} together, where the last two must also be summed over all allowed $n$, and premultiplying by $H(t)$, we obtain the solution to IBVP~\eqref{eq:nonnewt}:
\begin{equation}
u(y,t) = H(t)\left[(1 - y) - \frac{2}{\pi}\sum_{n=1}^\infty \exp\left(-\frac{1+\lambda_n\alpha}{2\tau}t\right) A_n(t) \frac{\sin(n\pi y)}{n}\right],
\label{eq:soln_ef_nonnewt_new}
\end{equation}
where
\begin{equation}
A_n(t) = 
\begin{cases}
\displaystyle\frac{1-\lambda_n\alpha}{\sqrt{\Delta_n}}\sinh\left(\displaystyle\frac{t}{2\tau}\sqrt{\Delta_n}\right) + \cosh\left(\displaystyle\frac{t}{2\tau}\sqrt{\Delta_n}\right), &\Delta_n > 0,\\[5mm]
\displaystyle\frac{1-\lambda_n\alpha}{2\tau}t + 1, &\Delta_n = 0,\\[5mm]
\displaystyle\frac{1-\lambda_n\alpha}{\sqrt{|\Delta_n|}}\sin\left(\displaystyle\frac{t}{2\tau}\sqrt{|\Delta_n|}\right) + \cos\left(\displaystyle\frac{t}{2\tau}\sqrt{|\Delta_n|}\right), &\Delta_n < 0.
\end{cases}
\label{eq:An_correct}
\end{equation}
Evidently the only difference between Eqs.~\eqref{eq:An_wrong} and \eqref{eq:An_correct} is that in the pre-factors of the $\sinh$, $t$ and $\sin$ terms $+\lambda_n\alpha$ has been changed to $-\lambda_n\alpha$.

Let us now try to resolve the apparent contradiction that Eqs.~\eqref{eq:soln_ef_nonnewt}--\eqref{eq:An_wrong} and  Eqs.~\eqref{eq:soln_ef_nonnewt_new}--\eqref{eq:An_correct}, which are not identical, were both found to be solutions to the same linear IBVP.

% ----------------------------------------------------------------
\section{Resolution to the apparent difficulty, or how to find the correct solution using an eigenfunction expansion}%Clarification on how to transform a start-up IBVP into an equivalent on with homogeneous  boundary conditions}
\label{sec:ef_correct}

It should be obvious that Eqs.~\eqref{eq:soln_ef_newt} and \eqref{eq:soln_ef_nonnewt} are not mathematically equivalent to Eqs.~\eqref{eq:soln_laplace_newt} and \eqref{eq:soln_ef_nonnewt_new} due to ``implicit assumptions'' about start-up. Only Eqs.~\eqref{eq:soln_laplace_newt} and \eqref{eq:soln_ef_nonnewt_new} account for all the stipulations of the start-up problem in a self-contained mathematical fashion. Thus, guided by these expressions, we modify the transformation in Eq.~\eqref{eq:subst_ss} as follows:
\begin{equation}
v(y,t) = u(y,t) - H(t)u_\mathrm{ss}(y).
\label{eq:subst_ss_Ht}
\end{equation}
Of course, one may argue this is merely a semantic point because it is understood that $u_\mathrm{ss}(y)$ in Eq.~\eqref{eq:subst_ss} only ``makes sense'' for $t>0$. However, because the governing PDEs contain partial time derivatives, the transformations in Eqs.~\eqref{eq:subst_ss} and \eqref{eq:subst_ss_Ht} are \emph{not} equivalent. Thus, it is important that the start-up condition is \emph{always} written \emph{explicitly} (though the appropriate $H(t)$ pre-factor) to obtain the physical solution to the start-up problem and, moreover, to not alter the state of rest prior to start-up.

Noting that $\partial \big(H(t)u_\mathrm{ss}(y)\big)/\partial t = \delta(t)u_\mathrm{ss}(y)$, where $\delta(t)$ is the Dirac delta distribution, and $\partial^2 \big(H(t)u_\mathrm{ss}(y)\big)/\partial y^2 = 0$, IBVP~\eqref{eq:newt} becomes
\begin{subequations}\begin{align}
\frac{\partial v}{\partial t} &= \frac{\partial^2 v}{\partial y^2} - \delta(t)u_\mathrm{ss}(y),\qquad (y,t) \in (0,1)\times(0,\infty),\label{eq:newt_subs_ss_pde_new}\displaybreak[3]\\
v(y,0) &= 0,\qquad 0<y<1,\displaybreak[3]\\
v(0,t) &= 0,\qquad t > 0,\\
v(1,t) &= 0,\qquad t > 0.
\end{align}\label{eq:newt_subts_ss_new}\end{subequations}
The initial condition, being understood as the state prior to start-up, i.e., as $\lim_{t\to0^-} u(y,t)$, naturally remains zero. This interpretation is a demonstration of \emph{Duhamel's principle} \citep{D33,BC42,S06}, namely, that a time-varying boundary condition can be ``exchanged'' for a homogeneous boundary condition at the ``cost'' of adding a time-varying source term to the linear BVP. Notice that the textbook approach exchanges the inhomogeneous boundary conditions for a homogeneous boundary condition at the cost of an inhomogeneous initial condition. Philosophically, this is already problematic because the cumulative effects of the boundary condition from $t=0$ up to $t=\infty$ have been ``condensed'' into an initial condition and imposed $t=0$, an act that readily violates the principle of causality, namely ``no output before the input'' \citep{T56}.\footnote{It should be noted that there are deeper issues regarding causality in relation to viscous (compressible) flow that are beyond the scope of the present work \citep{JMP00}.}

As before, the method of separation of variables suggests the ansatz $v(y,t) = \sum_n a_n(t)\psi_n(y)$. Substituting the latter into Eq.~\eqref{eq:newt_subs_ss_pde_new} and using the orthogonality relation from Eq.~\eqref{eq:eigenfunc}, we see that $a_n$ must satisfy
\begin{equation}
\frac{\rd a_n}{\rd t} = -\lambda_n a_n - \frac{2}{n\pi}\delta(t),\qquad a_n(0) = 0.
\end{equation}
Taking the Laplace transform, we obtain $(s+\lambda_n)\bar{a}_n = -2/(n\pi)$, and it follows that $a_n(t) = -2/(n\pi)H(t)\re^{-\lambda_n t}$.\footnote{Notice that the Laplace domain solution is identical to the corresponding one for the ODE in Eq.~\eqref{eq:ode_a_new}, where the $-2/(n\pi)$ term is contributed by the initial condition $a_n(0) = -2/(n\pi)$.} %{\color{red}See the Appendix for a longer discussion.}} 
Thus, the complete solution to IBVP~\eqref{eq:newt} is precisely
\begin{equation}
u(y,t) = H(t)\left[ (1 - y) - \frac{2}{\pi}\sum_{n=1}^\infty \exp\left(-n^2\pi^2 t\right)\frac{\sin(n\pi y)}{n}\right],
\label{eq:soln_ef_newt_new}
\end{equation}
which is identical to the Laplace-transform-in-time-only solution given in Eq.~\eqref{eq:soln_laplace_newt}. Unlike the expression in Eq.~\eqref{eq:soln_ef_newt}, the solution in Eq.~\eqref{eq:soln_ef_newt_new} is valid for all $t\in\mathbb{R}$, identically satisfying the condition that the fluid is at rest prior to $t=0$.

Similarly, IBVP~\eqref{eq:nonnewt} becomes
\begin{subequations}
\begin{multline}
\left(1 + \tau\frac{\partial}{\partial t}\right)\frac{\partial v}{\partial t} = \frac{\partial}{\partial y} \left(1 + \alpha\frac{\partial}{\partial t}\right)\frac{\partial v}{\partial y} - [\delta(t) + \tau\delta'(t)]u_\mathrm{ss}(y),\\ (y,t) \in (0,1)\times(0,\infty),\label{eq:nonnewt_subs_ss_pde_new}
\end{multline}
\vspace{-9mm}
\begin{align}
v(y,0) &= \frac{\partial v}{\partial y}(y,0) = 0,\qquad 0<y<1,\\
v(0,t) &= 0,\qquad t > 0,\\
v(1,t) &= 0,\qquad t > 0.
\end{align}\label{eq:nonnewt_subst_ss_new}\end{subequations}
Once again we note the effect of Duhamel's principle: the inhomogeneous boundary condition was exchanged for a source term.
Substituting the separation of variables ansatz into Eq.~\eqref{eq:nonnewt_subs_ss_pde_new} and using the orthogonality relation from Eq.~\eqref{eq:eigenfunc}, we find that $a_n$ satisfies
\begin{multline}
\left(1 + \tau\frac{\rd}{\rd t}\right)\frac{\rd a_n}{\rd t} = -\lambda_n\left(1 + \alpha\frac{\rd}{\rd t}\right) a_n - \frac{2}{n\pi}[\delta(t) + \tau\delta'(t)],\\ a_n(0) = \frac{\rd a_n}{\rd t}(0) = 0.
\end{multline}
Taking the Laplace transform of the latter, we obtain
\begin{equation}
(1+\tau s)s \bar{a}_n = -\lambda_n(1+\alpha s)\bar{a}_n-\frac{2}{n\pi}(1+\tau s).
\label{eq:a_ode_nonnewt_new_laplace}
\end{equation}
Meanwhile, if we take the Laplace transform of Eq.~\eqref{eq:a_ode_nonnewt}, we find that
\begin{equation}
(1+\tau s)s \bar{a}_n - (1+\tau s) a_n(0) - \tau \frac{\rd a_n}{\rd t}(0)  = -\lambda_n(1+\alpha s)\bar{a}_n + \lambda_n \alpha a_n(0),
\end{equation}
which, after imposing the initial conditions from Eq.~\eqref{eq:nonnewt_ef_a_ic}, becomes
\begin{equation}
(1+\tau s)s \bar{a}_n = -\lambda_n(1+\alpha s)\bar{a}_n - \frac{2}{n\pi} (1 + \tau s+ \lambda_n \alpha).
\label{eq:a_ode_nonnewt_new_laplace_wrong}
\end{equation}
Clearly, Equation~\eqref{eq:a_ode_nonnewt_new_laplace_wrong} does \emph{not} agree with Eq.~\eqref{eq:a_ode_nonnewt_new_laplace} unless $\alpha=0$ (a point on which we comment in Sect.~\ref{sec:discussion}).

\begin{figure}[!ht]
\centerline{\includegraphics[width=\columnwidth]{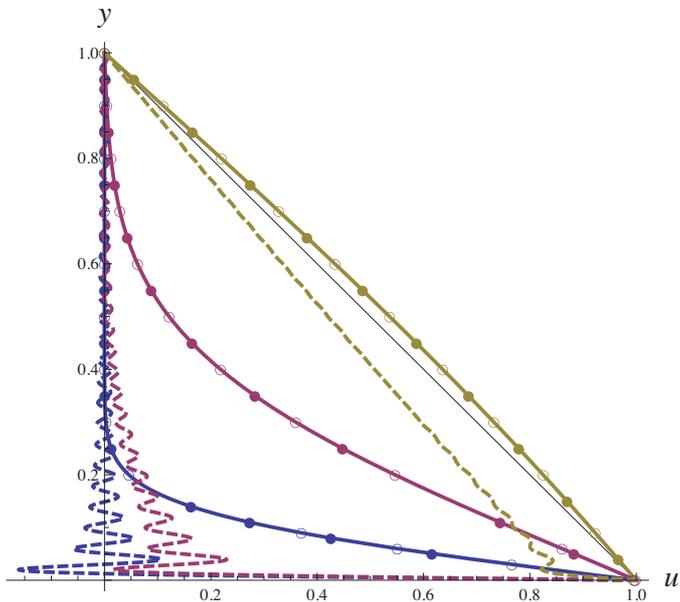}}
\caption{Approach to steady-state of the start-up flow of a Jeffreys' fluid (with $\tau = 1$ and $\alpha = 0.5$) between two parallel plates: 50 terms of Eqs.~\eqref{eq:soln_ef_nonnewt}--\eqref{eq:An_wrong} (bold dashed curves), 25 terms of Eqs.~\eqref{eq:soln_ef_nonnewt_new2}--\eqref{eq:An_correct2} (bold solid curves), Tzou's Riemann sum inversion from Eq.~\eqref{eq:Tzou} (open symbols), numerical solution of IBVP~\eqref{eq:nonnewt} (filled symbols) and the steady-state from Eq.~\eqref{eq:ss} (thin line). Darkest colors (blue online), intermediate dark colors (red online) and lightest colors (yellow online) correspond, respectively, to the solutions at $t=0.01$, $0.1$ and $1$.}
\label{fig:soln}
\end{figure}

Solving for $\bar{a}_n$ from Eq.~\eqref{eq:a_ode_nonnewt_new_laplace}, we find that
\begin{equation}
\bar{a}_n(s) = -\frac{2(1+\tau s)}{n\pi [(1+\tau s)s + \lambda_n(1+\alpha s)]} = -\frac{2(\tau^{-1} + s)}{n\pi (s-s^\bullet_+)(s-s^\bullet_-)}. %= -\frac{2}{n\pi \sqrt{\Delta_n}}\left(\frac{\sqrt{\Delta_n} + 1 + \tau s^\bullet_-}{s-s^\bullet_+} - \frac{1+\tau s^\bullet_-}{s-s^\bullet_-}\right).
\label{eq:a_bar_n_of_s_correct}
\end{equation}
where $s^\bullet_\pm = -\left[(1+\lambda_n\alpha)\pm\sqrt{\Delta_n}\right]/(2\tau)$ with $\Delta_n := (1+\lambda_n\alpha)^2 - 4\lambda_n\tau$. %, and the last equality only holds for $\Delta_n\ne0$. 
Using entries 11 and 18 from the Table of Laplace Transforms in \citep[pp.~317--321]{D74} for $\Delta_n\ne0$ and entries 8 and 17 for $\Delta_n = 0$, once again three cases can be distinguished
%\begin{equation}
%a_n(t) = -H(t)\frac{2}{n\pi}\exp\left(-\displaystyle\frac{1+\lambda_n\alpha}{2\tau}\right)\begin{cases}
%\displaystyle\frac{1-\lambda_n\alpha}{\sqrt{\Delta_n}}\sinh\left(\displaystyle\frac{t}{2\tau}\sqrt{\Delta_n}\right) + \cosh\left(\displaystyle\frac{t}{2\tau}\sqrt{\Delta_n}\right), &\Delta_n > 0\\[5mm]
%\displaystyle\frac{1-\lambda_n\alpha}{2\tau}t + 1, &\Delta_n = 0\\[5mm]
%\displaystyle\frac{1-\lambda_n\alpha}{\sqrt{|\Delta_n|}}\sin\left(\displaystyle\frac{t}{2\tau}\sqrt{|\Delta_n|}\right) + \cos\left(\displaystyle\frac{t}{2\tau}\sqrt{|\Delta_n|}\right), &\Delta_n < 0
%\end{cases}
%\end{equation}
and, finally, we arrive at the correct solution:
\begin{equation}
u(y,t) = H(t)\left[(1 - y) - \frac{2}{\pi}\sum_{n=1}^\infty \exp\left(-\frac{1+\lambda_n\alpha}{2\tau}t\right) A_n(t) \frac{\sin(n\pi y)}{n}\right],
\label{eq:soln_ef_nonnewt_new2}
\end{equation}
where
\begin{equation}
A_n(t) = 
\begin{cases}
\displaystyle\frac{1-\lambda_n\alpha}{\sqrt{\Delta_n}}\sinh\left(\displaystyle\frac{t}{2\tau}\sqrt{\Delta_n}\right) + \cosh\left(\displaystyle\frac{t}{2\tau}\sqrt{\Delta_n}\right), &\Delta_n > 0,\\[5mm]
\displaystyle\frac{1-\lambda_n\alpha}{2\tau}t + 1, &\Delta_n = 0,\\[5mm]
\displaystyle\frac{1-\lambda_n\alpha}{\sqrt{|\Delta_n|}}\sin\left(\displaystyle\frac{t}{2\tau}\sqrt{|\Delta_n|}\right) + \cos\left(\displaystyle\frac{t}{2\tau}\sqrt{|\Delta_n|}\right), &\Delta_n < 0.
\end{cases}
\label{eq:An_correct2}
\end{equation}
Clearly, this solution agrees exactly with the solution found in Sect.~\ref{sec:laplace} using only the Laplace transform in time, specifically Eqs.~\eqref{eq:An_correct} and \eqref{eq:An_correct2} are identical. Thus, though it may appear the differences between the Jeffreys-fluid solutions in Sect.~\ref{sec:ef_wrong} and this section are minor (almost typographical in nature), these two Fourier series representations have fundamentally different properties as illustrated in Fig.~\ref{fig:soln}.

One independent check on the exact solution for the start-up of plane Couette flow of a Jeffreys fluid found above is a direct numerical inversion of the Laplace-domain solutions (i.e., Eq.~\eqref{eq:soln_lap_dom_nonnewt}). This can be achieved by a Riemann-sum approximation \citep[Sect.~2.5.1]{T97}:
\begin{equation}
u(y,t) \approx \displaystyle{\frac{\re^{4.7}}{t}\left\{\frac{1}{2}\bar{u}\left(y,\frac{4.7}{t}\right)+\Real\left[ \sum_{m=1}^{M} (-1)^m \bar{u}\left(y,\frac{4.7+\ri m \pi}{t}\right)\right] \right\}},
\label{eq:Tzou}
\end{equation}
where the value $4.7$ has been fine-tuned numerically to ensure the accuracy of the approximation \citep[p.~41]{T97}, and $M\gg1$ is taken large enough to ensure that the sum has converged. Another independent check on the exact solution is a comparison to a numerical solution of IBVP~\eqref{eq:nonnewt} by the second-order-accurate finite-difference scheme constructed by \citet[Sect.~3.2]{C10} (using the same number of spatial and temporal grid points as therein). Both of these are shown as discrete symbols in Fig.~\ref{fig:soln}.

Since $\alpha$ and $\tau$ are dimensionless, for the purposes of Fig.~\ref{fig:soln} we take them to be $\mathcal{O}(1)$ without loss of generality. In accordance with experimental observations \citep{TS53}, we choose them such that $\alpha/\tau = 0.5$.

% ----------------------------------------------------------------
\section{Discussion}
\label{sec:discussion}

In this paper, we expounded on some subtle but critical aspects of the method of solution for transient/start-up problems in continuum mechanics. To the best of our knowledge, the erroneous approach of Sect.~\ref{sec:ef_wrong} was first introduced for the non-Newtonian case with $\tau=0$ by \citet{R82}. However, it did not become prominent in the literature until the paper by \citet{BR95} brought renewed interest to the topic (see also \citep{J05}). Meanwhile, the incorrect solution to the start-up plane Couette flow of a Jeffreys (nowadays referred to as ``Oldroyd-B'') fluid appears to have been supplied by \citet{HSA01,HKA04}.\footnote{It should be noted that \citet{HSA01,HKA04} do not present their solution technique, instead they only state the final form of the solution. Since \citep{R82} is cited in \citep{HSA01} and \citep{HSA01} is cited in \citep{HKA04}, we are led to believe that \citet{HSA01,HKA04} arrived at the wrong solution by the textbook technique (as in Sect.~\ref{sec:ef_wrong} herein).} As a result of many authors applying the erroneous technique from Sect.~\ref{sec:ef_wrong} to just about every variation of non-Newtonian rheology and boundary conditions, a vast literature of works presenting erroneous solutions now exists. For example, the incorrect solution for the start-up of plane Couette flow of a second-order fluid from \citep{R82} has made its way into textbooks \citep[Sect.~7.2]{TR00}.

Here, we do not attempt to catalogue the various erroneous solutions, including those in which the lower plate velocity is set into oscillatory or otherwise time-dependent motion such that $u(0,t) = H(t)\mathfrak{F}(t)$, in which case matters are complicated by the fact that the mathematical error is self-canceling if $\mathfrak{F}(0) = 0$. %Eq.~(7.2-34)
Detailed case studies by \citet{C10,C11} and \citet{CJ12} correct a related subset of the literature on non-Newtonian fluid mechanics in which erroneous solutions have been derived due to a (more mundane) mistake in applying the Fourier sine transform to a mixed derivative term (see also \citep{CJ09,CC10}). Furthermore, while the present work only deals with planar unidirectional flows, the same reasoning applies to start-up problems in cylindrical domains wherein a fluid (respectively, a heat-conducting material) fills a cylinder or the gap between two cylinders, one or both of which is rotating and/or translating (respectively, one or both of which is differentially heated). Indeed, \citet[Sect.~6]{BR95} commit the error from Sect.~\ref{sec:ef_wrong} for a cylindrical geometry and $\tau=0$, but proceed to (mistakenly) claim that the earlier solution to the same problem by \citet{R85} is wrong. Nevertheless, the approach of \citet{R85} is not rigorous because the non-zero value of the imposed boundary condition at $t=0$ has been set to zero. A careful derivation \citep{C11} shows that despite this arbitrary and unjustified assumption, one can obtain the correct solution (for the wrong reason).

For $\alpha=0$ both the erroneous (Eqs.~\eqref{eq:soln_ef_nonnewt}--\eqref{eq:An_wrong}) and correct (Eqs.~\eqref{eq:soln_ef_nonnewt_new2}--\eqref{eq:An_correct2}) solutions reduce to the correct solution to the problem of start-up of plane Couette flow of a Maxwell fluid \citep{DP71}.\footnote{The solution given by \citet{DP71} is indeed the correct one because it can also be obtained by using only the Laplace transform in time \citep{J11PC}. There are, however, two  typographical errors in \citep[Eq.~(8)]{DP71}: (i) the pre-factors $1/\lambda_n$ and $1/(-\lambda_n)$ of the $\sin$ and $\sinh$ terms, respectively, should be $1/\sqrt{\lambda_n}$ and $1/\sqrt{-\lambda_n}$, where $\lambda_n$ in \citep{DP71} corresponds to $\Delta_n$ in the present work, and (ii) $T_n$ should be multiplied by $-1$.} For $\tau = \alpha = 0$ the Newtonian solution is recovered in all cases. Of course, reducing one's solution to a published result in some limit is not a proof that the former is correct, claiming so would be a logical fallacy. The only conclusion that can be drawn is that terms of the solution that cannot be verified to be correct are multiplied by the parameter that is being taken to zero. Nevertheless, the pernicious fact that the even erroneous solutions reduce to correct ones in such limits seems to have deterred proper verification of some solutions in the literature.

At the heart of the problem presented in this paper is the matter of the \emph{start-up jump}.\footnote{It is important to note that this is distinct from the paradox identified by Sir Horace Lamb \citep{L27}, i.e., the discontinuity at $y=0$ of the function defined by the Fourier series $\sum_{n=1}^\infty \frac{1}{n} \sin(n\pi y)$, which is part of all the solutions presented herein in the limit as $t\to0^+$ (see also \citep[p.~191]{B67}). The way to reconcile both Lamb's paradox and the start-up jump phenomenon with our (perhaps unrealisitic) expectations of smoothness of solutions is to realize that the initial condition and the solution to the BVP for $t>0$ do not have to hold true at boundaries, as made explicit by the notation used to state IBVPs~\eqref{eq:newt} and \eqref{eq:nonnewt} herein. Indeed, the initial and boundary conditions for Stokes-type problems are \emph{incompatible}, leading to a discontinuity even at $t=0$.} This is the phenomenon in which the solution to a start-up problem is discontinuous across the plane $t=0$. It appears to have been first identified by Tanner \citep{T62a} in solving the classical first problem of Stokes for the Jeffreys fluid. In general, it is thought that start-up jumps are due to the presence of mixed derivatives in the governing partial differential equation, as confirmed by a number of examples in the literature most recently for start-up flows of dipolar fluids \citep[Eq.~(4.1)]{JP02} and bubbly liquids \citep{JF06}. However, the first exact solutions obtained by \citet{T63} for the second-order fluid ($\tau=0$) and by \citet{WK70} for the Jeffreys fluid did not exhibit start-up jumps, which appears to have led to the incorrect supposition in the literature that start-up jumps cannot occur for such non-Newtonian fluids \citep[see, e.g.,][p.~121]{TR00}.\footnote{Curiously, it appears that Eqs.~(9.25) and (9.28) of \citet{BR95}, which were obtained using only the Laplace transform in $t$, provide an alternative representation of the dimensional version of Eq.~(3.4) of \citet{J05}, i.e., the correct solution to IBVP~\eqref{eq:nonnewt} (with $\tau=0$). However, on the grounds that Eqs.~(9.25) and (9.28) of \citet{BR95} exhibit a start-up jump discontinuity, it was concluded by the latter authors that they do not represent the solution being sought.}

In conclusion, it is the author's recommendation that the textbook description of start-up problems of fluid flow and of transient heat conduction be revised to include the approach from Sect.~\ref{sec:ef_correct}, so that the treatment is mathematically rigorous and applicable to non-classical problems.

% ----------------------------------------------------------------
\section*{Acknowledgements}
This work was supported by NSF Grant DMS-1104047. The author is indebted to Dr.\ Pedro M.\ Jordan, Prof.\ Christo I.\ Christov, Prof. Howard A.\ Stone and Dr.\ Andrew J.\ Cihonski for constant encouragement and many helpful remarks and discussions both on the subject matter and on the manuscript.

\bibliographystyle{model2-names}
\bibliography{start-up}

% ----------------------------------------------------------------
\end{document}